# Colossal magnetocapacitance and colossal magnetoresistance in $HgCr_2S_4$

S. Weber[1], P. Lunkenheimer[1,*], R. Fichtl[1], J. Hemberger[1], V. Tsurkan[1,2], and A. Loidl[1]

[1]*Experimental Physics V, Center for Electronic Correlations and Magnetism, University of Augsburg, D-86135 Augsburg, Germany*
[2]*Institute of Applied Physics, Academy of Sciences of Moldova, MD-2028 Chisinau, R. Moldova*

We present a detailed study of the dielectric and charge transport properties of the antiferromagnetic cubic spinel $HgCr_2S_4$. Similar to the findings in ferromagnetic $CdCr_2S_4$, the dielectric constant of $HgCr_2S_4$ becomes strongly enhanced in the region below 60 – 80 K, which can be ascribed to polar relaxational dynamics triggered by the onset of ferromagnetic correlations. In addition, the observation of polarization hysteresis curves indicates the development of ferroelectric order below about 70 K. Moreover, our investigations in external magnetic fields up to 5 T reveal the simultaneous occurrence of magnetocapacitance and magnetoresistance of truly colossal magnitudes in this material.

The detection of colossal magnetoresistance in a number of perovskite-related manganites [1] maybe was the most notable discovery in solid state physics since the emergence of high-$T_c$ superconductors. Very recently another "colossal" effect with tremendous prospects for applications in modern microelectronics attracted considerable attention: In several, partly quite different materials an extremely strong coupling of magnetic and dielectric properties, termed "magnetocapacitive" or "magnetoelectric" effect, was found [2,3]. Among these, the cubic spinel $CdCr_2S_4$ stands out by showing simultaneous ferromagnetic (FM) and relaxor ferroelectric behavior [3] and by setting a record value of the magnetocapacitive effect with an increase of the dielectric constant *e'* of up to 3000% in a magnetic field of 10 T [4]. At the FM transition at $T_c \approx 84$ K, the dielectric constant shows a strong increase, which was ascribed to a speeding up of relaxational dynamics under the formation of magnetic order [4]. However, clearly the microscopic origin of the extraordinary behavior of $CdCr_2S_4$ is far from being understood. A similar, but weaker effect was recently reported in isostructural $CdCr_2Se_4$ [5] exhibiting FM order below $T_c \approx 125$ K. In the present letter, we provide dielectric data on $HgCr_2S_4$, which in contrast to $CdCr_2S_4$ does not show any long-range FM order but exhibits a complex antiferromagnetic (AFM) type of order below 22 K [6]. Nevertheless, it shows a colossal magnetocapacitance (CMC) of even larger amplitude than in $CdCr_2S_4$ and, most remarkably, the simultaneous occurrence of colossal magnetoresistance (CMR). In addition, our experiments indicate that in $HgCr_2S_4$, similar to our findings in $CdCr_2S_4$, short range ferroelectric order develops at low temperatures.

All measurements were performed on single crystals of $HgCr_2S_4$, grown by chemical transport. Measurements with silver paint and sputtered gold contacts applied to opposite sides of the samples were performed. Details on crystal preparation and experimental methods are given in Refs. [3,6]. A thorough characterization of the magnetic properties of $HgCr_2S_4$ [6] revealed an AFM state with non-collinear spin configuration below 22 K [7]. In moderate external magnetic fields up to 0.5 T, the AFM transition is shifted to lower temperatures and FM correlations, which are present already in zero field below 60 K [7,8], become strongly enhanced. At fields above 1 T, the AFM transition is completely suppressed and full FM order is induced at temperatures as high as 60 K. In Fig. 1(a) we show the magnetization curves for two temperatures. At 30 K, above the AFM transition, a typical soft-magnetic hysteresis curve is detected, which demonstrates how easily FM order is induced. At 5 K, in the AFM state, a clearly non-linear behavior is observed for low fields $H < 5$ kOe indicating a complex magnetic structure as discussed in detail in [6].

To check for possible ferroelectric correlations [3], electric polarization cycles were performed at different temperatures. As revealed by Fig. 1(b), at temperatures below about 70 K indeed hysteresis loops show up, becoming more pronounced with decreasing temperature. This finding indicates ferroelectric behavior, the absolute values of the polarization being significantly higher than in $CdCr_2S_4$ [3]. It was noted in [9] that ferroelectric-like hysteresis loops can also arise from non-intrinsic effects. Within this scenario, a decrease of the polarization with increasing frequency was predicted. Indeed such a behavior is observed in $HgCr_2S_4$ for frequencies between 1 Hz and 1 kHz (not shown). However, this effect can be ascribed to the intrinsic relaxational behavior documented in Fig. 2 (see discussion below) as the dipoles cannot follow the field for higher frequencies [10]. In [9], a second criterion for non-intrinsic behavior was given, namely the disappearance of the polarization saturation at high frequencies, which clearly is not fulfilled in $HgCr_2S_4$. To further corroborate ferroelectric ordering in this system, we tried to measure the pyro current in order to obtain information on the temperature-dependent polarization. Unfortunately, these experiments were hampered by the relatively high conductivity of the samples at low temperatures. Overall, while the results of

Fig. 1(b) strongly suggest ferroelectric ordering in HgCr$_2$S$_4$, for a final proof still further experiments are necessary.

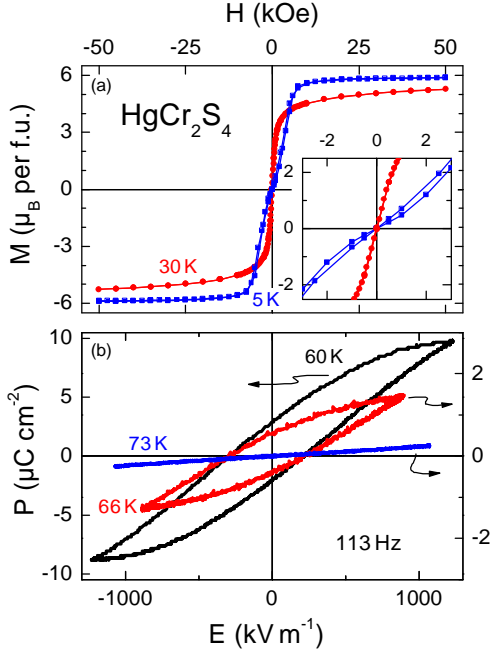

FIG. 1. (a) Magnetization hysteresis curves for two temperatures. The inset gives a magnified view of the behavior at small fields. (b) Polarization hysteresis curves for three temperatures measured at 113 Hz.

In Fig. 2 the temperature-dependent dielectric constant $\varepsilon'$ (a) and the conductivity $\sigma'$ (b) are shown for various frequencies at temperatures below 150 K. With decreasing temperature, $\varepsilon'(T)$ exhibits a pronounced steplike increase in the region between 60 and 80 K, which is followed by a decrease at low temperatures. Concomitantly, the conductivity reveals peaks at frequencies coinciding with the points of inflection of $\varepsilon'(T)$ [indicated by arrows in Fig. 2(b)]. They are superimposed to a frequency independent contribution, observed for the lower frequencies, with a peak at about 25 K, which can be ascribed to dc charge transport. Except for the dc contribution, which is much higher in the present case, qualitatively this behavior resembles that reported in CdCr$_2$S$_4$ [3,4]. However, while in CdCr$_2$S$_4$ an increase of $\varepsilon'(T)$ by about a factor of five was observed, in HgCr$_2$S$_4$ it extends over nearly two decades reaching values of $\varepsilon'$ as high as 2000! The finding of a strong frequency dependence of the step/peak frequencies indicates a relaxational origin of this phenomenon. For relaxational processes, steps in $\varepsilon'(T)$ and peaks in $\sigma'(T)$ (or the loss $\varepsilon''(T) \sim \sigma'(T) / \nu$) occur when the frequency of the exciting field is equal to the temperature-dependent relaxation rate of the relaxing entities, e.g., reorienting dipoles or charged particles jumping within a double-well potential. Obviously, with decreasing temperature in HgCr$_2$S$_4$ this condition is fulfilled twice, namely in the regions between 40 K and 70 K and again below about 20 K. In the latter region, the peak temperature decreases with decreasing frequency, mirroring the slowing down of the relaxational dynamics for low temperatures. However, between 40 K and 70 K the peaks shift in an opposite way, which, in line with the interpretation of the results in CdCr$_2$S$_4$ [4], indicates an unusual *acceleration* of the dynamics with decreasing temperature. It is this increase of the relaxational mobility that leads to the strong upturn of $\varepsilon'$ below about 60 - 80 K. We investigated a number of different samples, using contacts prepared by sputtered gold or silver paint and in all cases a behavior, very similar to that documented in Fig. 2 was detected. While above about 150 K contributions from electrode polarization [11] prevent the detection of the intrinsic properties of the samples, the results with different contact types reveal that for lower temperatures any contact contributions can be excluded and the detected relaxational behavior can be regarded as truly intrinsic. However, we found a strong variation of the dc conductivity for different measurements, which seems to strongly depend on purity, defects, and thermal history of the samples. Concerning the origin of the polar moments, currently only speculations are possible, e.g. about an off-center position of the Cr$^{3+}$-ions [3], similar to the situation in perovskite ferroelectrics (e.g., BaTiO$_3$), an intrinsic surface-related Maxwell-Wagner mechanism [4,12], or more exotic scenarios as, e.g., electronic ferroelectricity driven by charge order and/or coupling of the electron system to the lattice [13,14]. For $CdCr_2S_4$, the occurrence of ferroelectricity due to the softening of a polar mode has been excluded in recent LSDA+$U$ calculations [15].

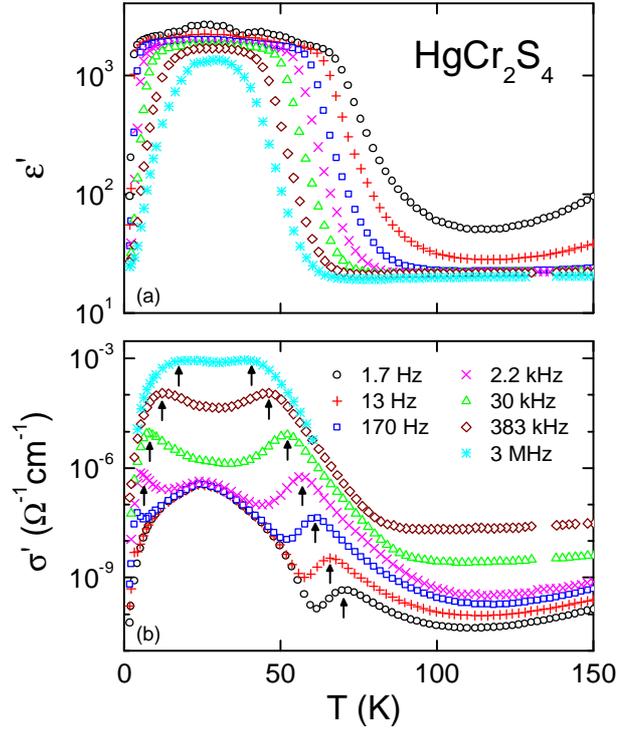

FIG. 2. Temperature dependence of dielectric constant (a) and conductivity (b) of HgCr$_2$S$_4$ for various frequencies. The arrows in (b) indicate the frequency-dependent position of the relaxation peaks.

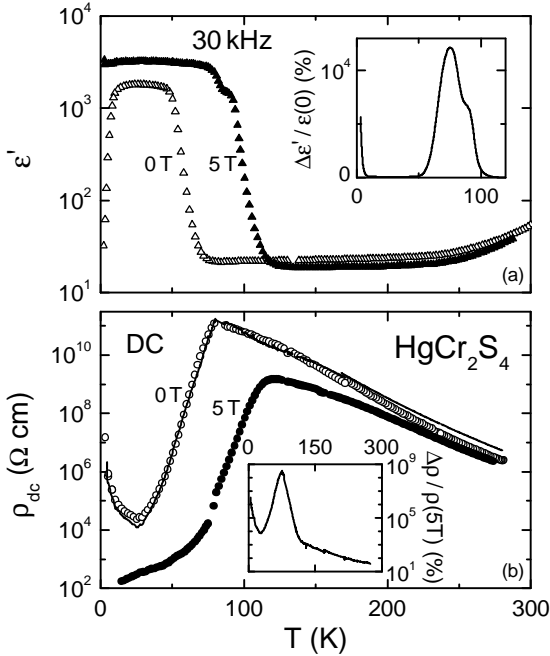

FIG. 3. (a) Temperature dependence of the dielectric constant at 30 kHz measured at zero field and in an external magnetic field of 5 T. The inset provides a measure of the magneto-capacitive effect, with $\Delta\varepsilon' = \varepsilon'(5T) - \varepsilon'(0T)$. (b) Temperature dependence of the dc resistivity at zero field and in a magnetic field of 5 T measured in two-point contact configuration (symbols). In addition, the results of a four-point measurement at zero field is shown (line), scaled by a multiplicative factor to match the two-point result. The inset shows the relative change of the resistivity with $\Delta\rho = \rho(0T) - \rho(5T)$.

It is suggestive to assume a connection of the anomaly in $\varepsilon'(T)$ at 60 – 80 K with the observation of FM correlations at 60 K [6,7,8]. Thus, as these correlations become strongly enhanced already in moderate external magnetic fields, in $HgCr_2S_4$ a considerable magnetocapacitance may be expected. In Fig. 3(a), we show the dielectric constant measured at 30 kHz without field and with a magnetic field of 5 T. The field indeed induces a strong shift of the upturn of $\varepsilon'$ of about 50 K towards higher temperatures. This corresponds to a huge field-dependent increase of $\varepsilon'$ in the region of 60 – 100 K. Similar behavior is also found at other measuring frequencies. As revealed by the inset of Fig. 3(a), the magnetocapacitance in $HgCr_2S_4$, defined as relative change of $\varepsilon'$, reaches values as high as 12000 %. In Fig. 4, the frequency dependence of $\varepsilon'$ measured for different magnetic fields is shown revealing a step-like decrease with increasing frequency, again typical for a relaxational process. As the frequency of the point of inflection gives an estimate of the relaxation rate, its shifting to higher frequencies with increasing magnetic field gives evidence for an acceleration of the relaxational dynamics by the magnetic field. The inset of Fig. 4 demonstrates the strong increase of the dielectric constant in external magnetic fields, which critically depends on the measuring frequency. It is remarkable that, in contrast to the findings in $CdCr_2S_4$ [3], $\varepsilon'(T)$ at 5 T [Fig. 3(a)] does not show a final reduction at low temperatures due to the slowing down of relaxational dynamics. Instead, down to the lowest temperature investigated (2.5 K), $\varepsilon'(T)$ remains constant. One may speculate about tunneling processes playing a role here and further experiments at lower temperatures are necessary to clarify this issue.

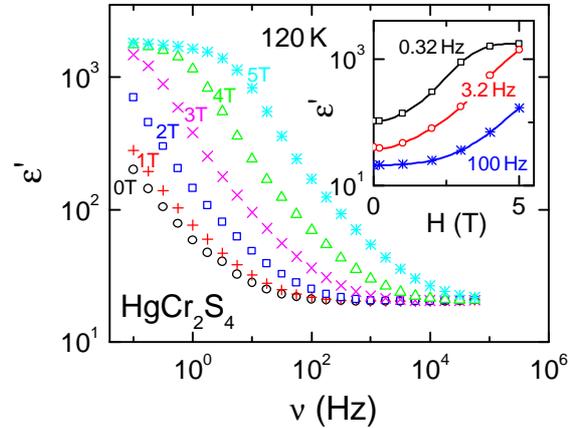

FIG. 4. Frequency-dependent dielectric constant at 120 K for various external magnetic fields between 0 and 5 T. The inset shows the field dependence for three selected frequencies.

In Fig. 3(b), the dc resistivity of $HgCr_2S_4$, measured using two-point (open circles) and four point contact geometry (line) is shown. Due to the small dimensions of the sample it was not possible to prepare contacts of well-defined geometry for the four-point measurement. As thus there is a high uncertainty in the absolute values of the resistivity in four-point geometry, the resulting curve was scaled by a multiplicative factor to match the two-point curve leading to a good agreement. Since a contact resistance only could give rise to an *additive* contribution, which in addition should be temperature dependent, the good match of the scaled two-point and the four-point curve gives further evidence that contact contributions can be neglected. $\rho_{dc}(T)$ exhibits a strong anomaly at about 80 K, switching from semiconducting to metal-like temperature characteristics. Finally the semiconducting behavior is restored below 25 K when the system orders antiferromagnetically. In an external magnetic field of 5 T, this anomaly is shifted to about 120 K, and the metal-like behavior is retained down to the lowest temperatures.

This corresponds to a strong variation of the resistivity with magnetic field; e.g., at 80 K $r_{dc}$ becomes reduced by about six decades in a field of 5 T. As shown in the inset of Fig. 3(b), the relative variation of $r_{dc}$ exceeds $10^8$ %, a truly colossal effect exceeding the findings in most CMR manganites [1]. Thus, obviously the CMC in HgCr$_2$S$_4$ is accompanied by the occurrence of CMR. The presence of FM correlations at low temperatures can be assumed to lead to the observed reduction of the resistivity in zero field. The enhancement of these correlations in an external magnetic field induces the detected magnetoresistance. In contrast to the CMR manganites [1], double exchange can be excluded in the present case, Cr$^{3+}$ having a half-filled $t_{2g}$ level, and the observed CMR probably results from the reduction of spin-disorder scattering in the induced FM state, an effect that may be considerably enhanced for low charge carrier densities [16].

We want to emphasize that the dc conductivity and the dielectric constant are completely independent quantities and an anomaly of one of those quantities not necessarily has to lead to an anomaly of the other. While the dielectric constant is related to the conductivity via the Kramers-Kronig relation, only ac conductivity, e.g. arising from hopping conduction of localized charge carriers [17], but not dc conductivity, can induce a corresponding contribution in $e'$. Ac conductivity, usually following a fractional power law $s' \propto n^s$ with an exponent $s < 1$, leads to a divergence of the dielectric constant for decreasing frequency [17,18]. Thus the saturation at a value of about 2000, observed for low frequencies and high fields in Fig. 4, clearly excludes ac conductivity as an explanation of the observed CMC. However, from a theoretical point of view, e.g., based on the Clausius-Mosotti equation or other theoretical considerations [19,20], a simultaneous decrease of the dc resistivity and increase of the dielectric constant can arise when approaching a metal-insulator transition from the insulating side. Indeed such a behavior was observed in some doped semiconductors [19,21]. However, these predictions concern purely electronic contributions to the dielectric constant, which should not show any notable frequency dispersion in the frequency range investigated in the present work. Nevertheless, it cannot be excluded that the electron system contributes to the observed behavior via a coupling to atomic displacements as discussed, e.g., in [13].

In summary, the spinel compound HgCr$_2$S$_4$ shows the simultaneous occurrence of CMC and CMR, both being of exceptionally large magnitude. Especially, despite this material exhibits no long-range FM order, its magnetocapacitance exceeds by far that observed in the FM spinel compounds CdCr$_2$S$_4$ [3,4] and CdCr$_2$Se$_4$ [5]. The reason for the exceptional behavior of HgCr$_2$S$_4$ must be found in the strong magnetic frustration as revealed by magnetic and specific-heat studies [6]. It leads to a complex AFM ground state, which can be easily switched into a FM state by moderate fields. Interestingly, structural investigations suggest that the system is close to a structural instability and that there is a possible off-centre displacement of the Cr$^{3+}$-ions [6]. It was speculated already in [3] that such an off-center position, in conjunction with geometrical frustration, may lead to a relaxor ferroelectric state and that the coupling of the relaxational dynamics to the magnetic order parameter may arise from a softening of the lattice via exchangestriction. But also other scenarios seem possible and further work has to be done to clarify the origin of the amazing magnetoelectrical properties of the chalcogenide-chromium spinels.

This work was partly supported by the Deutsche Forschungsgemeinschaft via the Sonderforschungsbereich 484 and partly by the BMBF via VDI/EKM.